\documentclass{elsart}

\usepackage{natbib}

\usepackage{graphicx}
\usepackage{amssymb}

\begin{document}

\begin{frontmatter}



\title{Metal Abundances in the ICM as a Diagnostics of the Cluster History}


\author{H. B\"ohringer$^1$, K. Matsushita$^2$, A. Finoguenov$^1$, Y. Xue$^3$, E. Churazov$^4$}

\address{$^1$ Max-Planck-Institut f\"ur extraterrestrische Physik, D-85748
Garching, Germany, $^2$~Dept. of Physics, Tokyo University of Science, Tokyo 162-8601, Japan,
$^3$~National Astronomical Observatories, CAS, Beijing
100012, China, $^4$~Max-Planck-Institut f\"ur Astrophysik, D-85748 Garching, Germany}

\begin{abstract}
Galaxy clusters with a dense cooling core exhibit a central increase
in the metallicity of the intracluster medium. Recent XMM-Newton
studies with detailed results on the relative abundances of 
several heavy elements show that the high central abundances 
are mostly due to the contribution from supernovae type Ia. The dominant
source is the stellar population of the central cluster galaxy. 

With this identification of the origin of heavy elements and the observed
rates of SN Ia in elliptical galaxies, the central abundance peak can be used
as a diagnostic for the history of the cluster core region. We find for four
nearby cooling core clusters that the enrichment times for the central peaks
are larger than 6 - 10 Gyrs even for a higher SN Ia rate in the past. 
This points to an old age and a relatively quiet history of these cluster 
core regions.

A detailed analysis of the element abundance ratios provides evidence that the
SN Ia yields in the central cluster galaxies are more rich in intermediate
mass elements, like Si and S, compared to the SN Ia models used to 
explain the heavy element enrichment in our Galaxy.   
\end{abstract}


\end{frontmatter}

% main text
\section{Introduction}
\label{}

Our understanding of the complex physics prevailing in cooling cores in 
clusters of galaxies is mostly driven by observation rather than by theory.
It was for example the large amount of X-ray radiation detected from
cluster cores that led to the scenario of cooling flows 
(e.g. Fabian \& Nulsen 1977, Fabian 1994). 
And it was later again observations by XMM-Newton showing that no steadily
cooling gas is observed and by Chandra displaying interaction effects
of central AGN with the cooling core plasma which lead to major revision
of the physical model of cluster cooling cores with much smaller mass deposition
rates than inferred previously (e.g. Peterson et al. 2001,  Matsushita et al. 2002,
McNamara et al. 2000, Forman et al. 2004,  Churazov et al. 2000, 2001,
B\"ohringer et al. 2002).  
To understand in detail the interaction effects of the central AGN with the
ICM and how the ICM is heated in a globally homogeneous and fine tuned way is
even more complicated. We have to hope again to exploit further observational
effects to guide our interpretation. In this context the entropy distribution
and the distribution of heavy elements in the ICM provide another important
diagnostic tool to learn more about the energy recycling and transport
processes in the central ICM. This conference contribution will concentrate on the
abundance distributions as diagnostics.

\begin{figure}
\centerline{\includegraphics[clip=,width=0.70\textwidth]{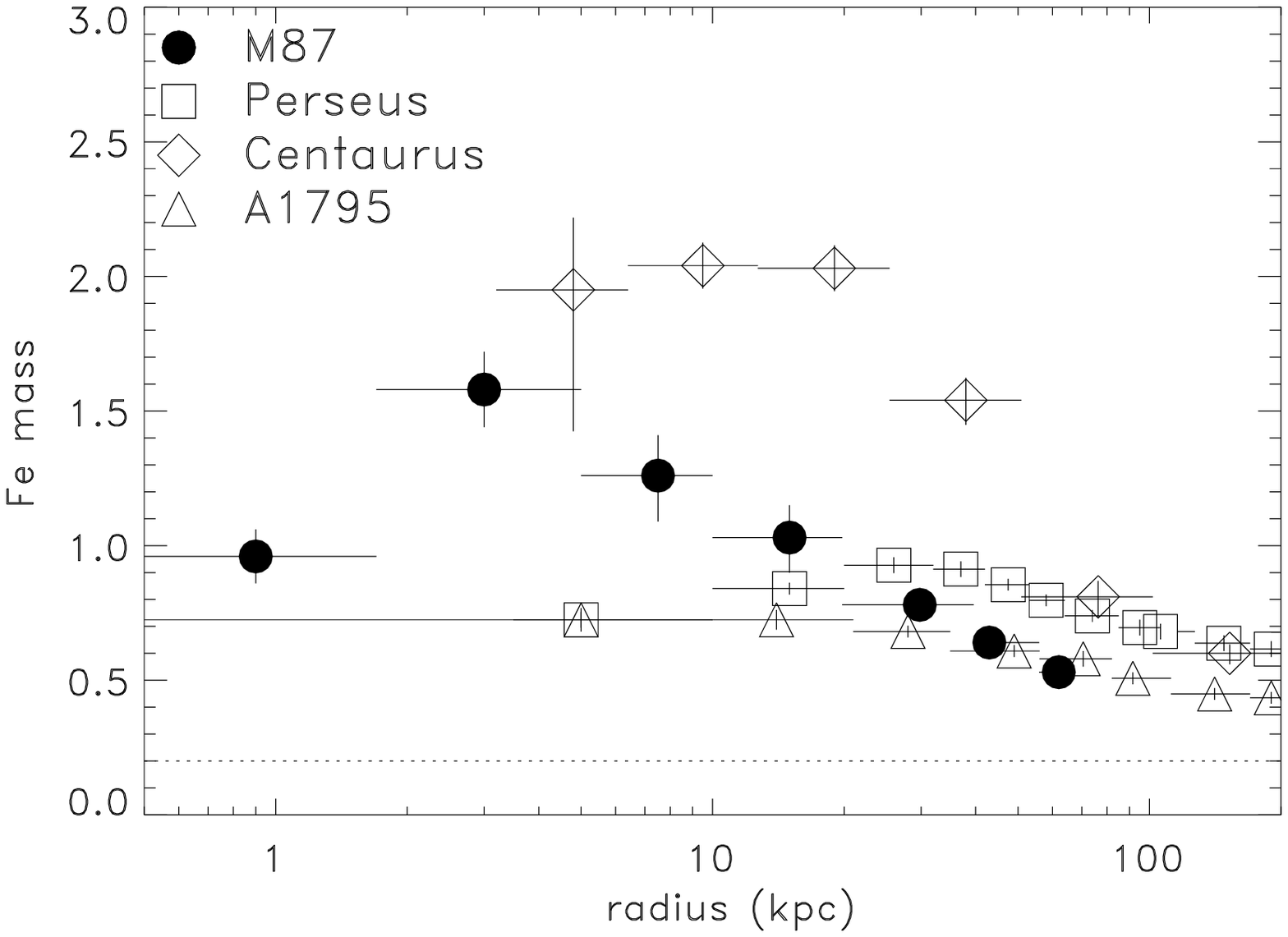}}
\caption{Relative abundance of Fe in the ICM as a function of cluster
radius in four nearby clusters (identified by the four different symbols) 
as deduced from XMM-Newton observations (B\"ohringer et al. 2004).
The dashed line shows the Fe abundance value of 0.2 solar, which is observed
at large radii in clusters.}
\label{Fig1}
\end{figure}

To model the observed abundance distribution of heavy
elements we need to know three things: the sources, the time scales, and the 
transport processes that redistribute the ICM.     

\section{The origin of the observed abundance distributions}
\label{}

\begin{figure}
\centerline{\includegraphics[clip=,width=0.98\textwidth]{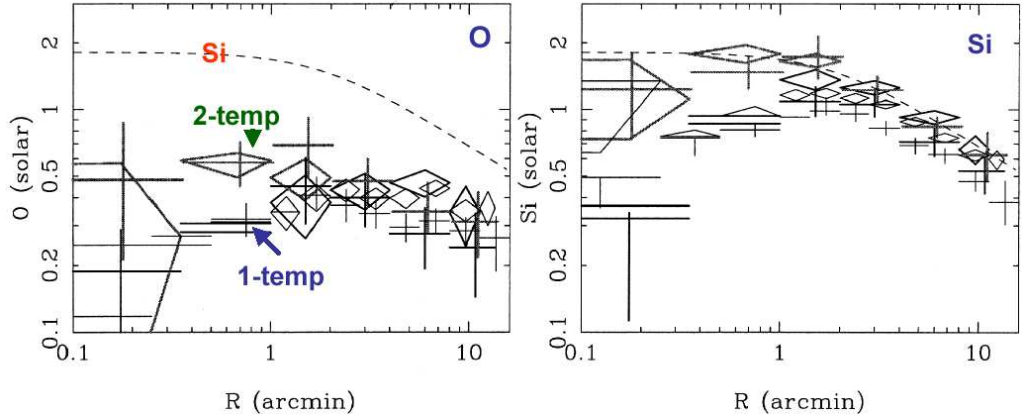}}
\caption{Abundance profiles of O and Si in the X-ray halo of M87 in the Virgo
  cluster as deduced from XMM-Newton observations (Matsushita et al. 2003). The upper 
curves for the two-temperature model give the more realistic results.}
\label{Fig2}
\end{figure}

We know two sources for the enrichment of the ICM with heavy elements: 
core collapse supernovae, type II, which produce a broad spectrum of element
masses with a bias towards the lighter elements like O and Mg, and type Ia
supernovae, thermonuclear explosions of white dwarfs, which dominantly
yield Fe group elements and lighter elements like Si and S but very little O
and Mg. One of the characteristics of cooling core clusters is the central
abundance peak of Fe and the heavier elements (e.g. DeGrandi \& Molendi 2001) 
as shown in Fig. 1. A more detailed inspection which elements follow this
abundance peak shows that it is traced by the heavier SN Ia products but not by
the lighter elements like O and Mg (Fig.2, Matsushita et al. 2003). This 
is consistent with the picture that SN II activity happens in the early
history of cluster formation when the stellar populations of the cluster
galaxies are still young. These elements have time to mix well in the ICM and
show a more homogeneous distribution. SN Ia are still occuring and observed in
the present day cluster ellipticals and in the central cD galaxies. The more
recent yields obviously lead to more local enrichments. In particular the
massive stellar population of the cD galaxies dominating the centers of cooling
core clusters are responsible for the central enrichment. 

\begin{figure}
\centerline{\includegraphics[clip=,width=0.70\textwidth]{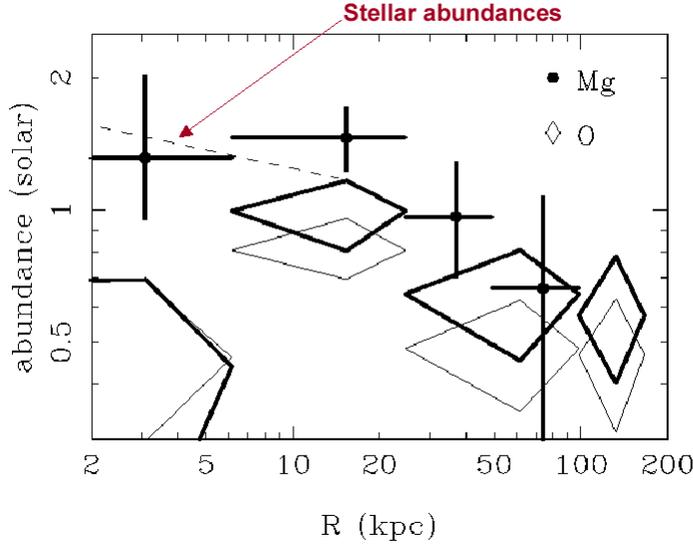}}
\caption{Relatve abundance of O and Mg in the central ICM of the
Centaurus cluster observed with XMM-Newton (Matsushita et al. 2004). The dashed line
shows the stellar Mg abundance for comparison.}
\label{Fig3}
\end{figure}

An even more careful inspection of the central abundances reveals also slight
increases in the relative abundances of O and Mg as shown in Fig. 3 for the 
case of the Centaurus cluster (Matsushita et al. 2004). The ICM abundances of
these elements are comparable to the stellar abundances which implies that the
ICM in the very central region most probably reflects the stellar mass loss of
the central galaxy. Consideration of the budget of the central ICM for example
for the case of M87 shows that
the inner 10 kpc region of the ICM, roughly the region of the scale radius
of cD galaxies (or approximately half light radius), contains a gas mass of
about $2\cdot 10^9$ M$_{\odot}$ and an iron mass of about 
$6\cdot 10^6$ M$_{\odot}$. Both can be replenished by stellar mass loss from
M87 in about 2 - 3 Gyrs
if we adopt a stellar mass loss rate of $2.5 \cdot 10^{-11} \times L_B$
(Ciotti et al. 1991) and a Fe production with a supernova rate of 0.15 SNU
(Cappelaro et al. 1999). It is important, however, for this enrichment to
accumulate that the gas is not cooling and condensing with the
classically calculated cooling flow rates, since in this case the central ICM
produced by stellar mass loss would disappear faster than it can be
replenished. Both, the evidence described in the previous section and the
observed abundances of O and Mg support the notion that the central ICM is not
condensing at the high rates inferred previously within cooling flow models.

\begin{figure}
\centerline{\includegraphics[clip=,width=0.98\textwidth]{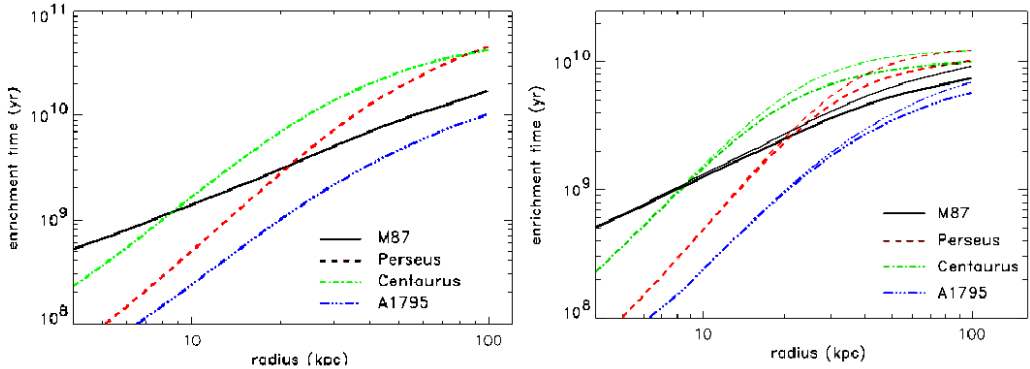}}
\caption{Enrichment ages of the central iron abundance peak
in the four galaxy clusters, M87/Virgo, Centaurus,
Perseus, and A1795. The enrichment times were calculated
on the basis of a SN Ia rate of 0.15 SNU
(Capellaro et al. 1999) and an additional contribution by stellar mass loss.
{\bf Left panel:} with constant SN Ia rate, {\bf Right panel:} with increasing 
SN Ia rate in the past described by a power law with time 
exponents $s=-2$ (thick lines) and $s=-1.1$. (thin lines).}
\label{Fig3}
\end{figure}

\section{Enrichment times}
The central abundance peak extends much further than the central 10 kpc region
inside the central galaxy. Since the central galaxy dominates the stellar
population out to radii of the order of 100 kpc and is therefore responsible
for most of the secular heavy elements in this zone, this implies some transport
of the centrally enriched ICM to larger radii and larger enrichment times than
inferred above for the 10 kpc region. In Fig. 4 we show the enrichment times
for the Fe abundance peak as a function of radius. The enrichment times were 
calculated from the total amount of excess Fe inside a given radius divided by the
iron yield inferred for the integrated stellar light of the cD galaxy with
the above quoted Ciotto et al. rate for stellar mass loss and the Cappelaro et al.
rate supernovae type Ia with a Fe mass yield per supernove of 0.7 M$_{\odot}$
(for more details of these calculations see B\"ohringer et al. 2004).
The excess Fe abundance is thereby determined by subtracting an ubiquitous Fe
abundance of  0.15 solar believed to originate from early SII enrichment.
Alternatively we were also considering the likely case that the type Ia supernova
rate was higher in the past and varied with time in a power law behavior. 
Even for this case, where the enrichment times are shortened, we find quite large
enrichment times of $5 - 9$ Gyrs for radii around 50 kpc and $7 - 12$ Gyrs 
for 100 kpc. 

If these estimates are correct the results imply that the central regions of
the clusters and their ICM can not have suffered major disturbances during
these times and that the ICM experienced only mild turbulent redistribution
as signified by the abundance peak which is broader than the light distribution
of the central galaxy. The central regions of cooling flow clusters must be
old with ages of the order of 10 Gyrs! Again, this scenario does not allow
the inclusion of cooling flows with the large previously inferred mass
deposition rates, since this would make the accumulation of the large observed 
central Fe mass excess of the order of $10^9$ M$_{\odot}$ impossible.
See also De Grandi et al. (2004) for a complementary discussion of the 
origin of the central Fe abundance from the SN yields integrated over the
entire past history.

\section{Supernova yields}
Adopting this scenario for the enrichment of the central heavy element
abundance peak, we can now study the element abundance ratio in more detail
and compare the results with predictions of the supernova nucleosynthesis
model calculations. The best measured abundances for the different 
atomic mass groups are those of Fe, Si 
and O. Thus we will concentrate here on the use
of these elements.   

\begin{figure}
\centerline{\includegraphics[clip=,width=0.84\textwidth]{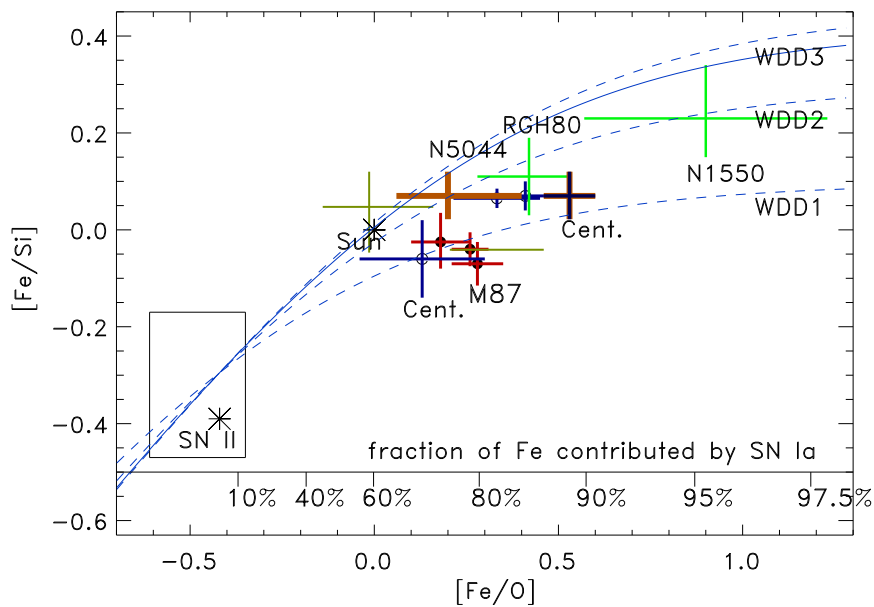}}
\caption{Comparison of the Fe/O and Fe/Si abundance ratios for different 
radial regions (with $r \le 100$ kpc, except for the left light green data point
for low temperature clusters for $r = 50 - 200$ kpc)
in different galaxy clusters with data taken from Matsushita et al. 
(2003 - M87; red), Matsushita et al. (2004 - Centaurus, blue), Sun et al. (2003 - NGC1550, green),
Xue et al. (204 - RGH80, green), Buote et al. (2003 - NGC5044, orange thick), Tamura et al. (2004 -
several cool clusters observed with XMM, light green thin), and Dupke \& White (2000 - A496
observed with ASCA, blue hidden behind N5044 data points). 
The solid line gives the W7 model (Nomoto et al. 1984) which 
accounts for the abundance of our sun. The three dashed lines give the 
delayed deflagration models WDD1, WDD2, and WDD3 by Iwamoto et al.(1999).
The lower asterisk indicates the SN II yields, the upper asterisk the composition
of the Sun. The box indicates the composition of low metallicity halo stars in
our galaxy (from Clementi et al. 1999).
The range of observational values is well bracketed by the 
different theoretical SN Ia yields.}
\label{Fig3}
\end{figure}

In Fig. 5 we compare the abundance ratios of Fe/O and Fe/Si in different regions
in different groups and clusters of galaxies. In this diagram 
the Fe/O ratio is a diagnostics of the ratio
of Fe yields from SN II and SNIa and the Fe/Si ratio can then be used to test
various models of SN Ia yields (Matsushita et al. 2003). 
In the plot we have compiled the available set of
observational results from XMM and Chandra (and one ASCA result) from the literature
and compare them to various models with different deflagration speeds of the
SN Ia explosion from Nomoto et al. (1984) and Iwamoto et al. (1999).
The models lines were calculated by mixing an increasing amount of SN Ia
material with the early composition taken to be similar to the average
of the low metallicity galactic halo stars (Clementini et al. 1999) which is also
close to the expected SN II yields as shown in the figure.
The results show that the bulk of the data lay below the curve of the W7 model
which fits the solar abundance and is successfully used to describe the
chemical history of our galaxy (Edvardsson et al. 1993). The WDD1 and WDD2 models 
provide yields with larger Si mass fractions compared to iron 
than supplied by the W7 model. The data points for M87 are
the most extreme in requiring a relatively large Si yield (Matsushita et
al. 2003). The possible reason for this is a more incomplete nuclear burning
in SN Ia explosions in older stellar populations (Finoguenov et al. 2001)
which is also consistent with the observed statistically lower luminosity of
SN Ia lightcurves observed in supernovae type Ia in elliptical galaxies
compared to spirals (Ivanov et al. 2000). 

\section{Conclusions} The results presented from the analysis of a few nearby
clusters studied with XMM-Newton and CHANDRA provide new clues on the
dynamical history of the central regions of clusters, the transport processes
in the ICM, and the nucleosynthesis yields of the two main types of
supernovae. The hopefully long life of the XMM and CHANDRA observatories will
allow us to extend these studies systematically to a well selected larger
sample of groups and clusters of galaxies, which will finally show if the
conclusions drawn in this contribution can be generalized and established with
higher precision.

\section{Acknowledgements}

The majority of the results have been obtained with 
XMM-Newton, an ESA mission with instruments and contributions directly funded
by ESA member States and the USA (NASA). The XMM-Newton Project is supported 
by the Bundesministerium f\"ur Bildung und Forschung, the Max-Planck-Society,
and the Haidenhain Stiftung. 

% Parenthetical: \citep{Bai92} produces (Bailyn 1992).
% Textual: \citet{Bai95} produces Bailyn et al. (1995).
% An affix and part of a reference:
%   \citep[e.g.][Ch. 2]{Bar76}
%   produces (e.g. Barnes et al. 1976, Ch. 2).

\end{document}